\documentclass[PPCF, 10pt]{iopart}
\usepackage{graphicx}
\usepackage{amssymb}
\usepackage[T2A]{fontenc}
\usepackage[utf8]{inputenc}
\usepackage[russian,english]{babel}
\usepackage{soul}
\usepackage{hyperref}
\hypersetup{
  colorlinks=true,
  allcolors=blue
}
\usepackage{xparse}
\relax

\DeclareDocumentCommand\vectorunit{ s m }{\IfBooleanTF{#1}{\boldsymbol{\hat{#2}}}{\mathbf{\hat{#2}}}} 
\DeclareDocumentCommand\vu{}{\vectorunit}

\DeclareDocumentCommand\vectorbold{ s m }{\IfBooleanTF{#1}{\boldsymbol{#2}}{\mathbf{#2}}} 
\DeclareDocumentCommand\vb{}{\vectorbold} 

\DeclareDocumentCommand\differential{ o g d() }{ 
	\IfNoValueTF{#2}{
		\IfNoValueTF{#3}
			{\diffd\IfNoValueTF{#1}{}{^{#1}}}
			{\mathinner{\diffd\IfNoValueTF{#1}{}{^{#1}}\argopen(#3\argclose)}}
		}
		{\mathinner{\diffd\IfNoValueTF{#1}{}{^{#1}}#2} \IfNoValueTF{#3}{}{(#3)}}
	}
\DeclareDocumentCommand\derivative{ s o m g d() }
{ 
	\IfBooleanTF{#1}
	{\let\fractype\flatfrac}
	{\let\fractype\frac}
	\IfNoValueTF{#4}
	{
		\IfNoValueTF{#5}
		{\fractype{\diffd \IfNoValueTF{#2}{}{^{#2}}}{\diffd #3\IfNoValueTF{#2}{}{^{#2}}}}
		{\fractype{\diffd \IfNoValueTF{#2}{}{^{#2}}}{\diffd #3\IfNoValueTF{#2}{}{^{#2}}} \argopen(#5\argclose)}
	}
	{\fractype{\diffd \IfNoValueTF{#2}{}{^{#2}} #3}{\diffd #4\IfNoValueTF{#2}{}{^{#2}}}}
}
\DeclareDocumentCommand\dv{}{\derivative} 
\DeclareDocumentCommand\partialderivative{ s o m g g d() }
{ 
	\IfBooleanTF{#1}
	{\let\fractype\flatfrac}
	{\let\fractype\frac}
	\IfNoValueTF{#4}
	{
		\IfNoValueTF{#6}
		{\fractype{\partial \IfNoValueTF{#2}{}{^{#2}}}{\partial #3\IfNoValueTF{#2}{}{^{#2}}}}
		{\fractype{\partial \IfNoValueTF{#2}{}{^{#2}}}{\partial #3\IfNoValueTF{#2}{}{^{#2}}} \argopen(#6\argclose)}
	}
	{
		\IfNoValueTF{#5}
		{\fractype{\partial \IfNoValueTF{#2}{}{^{#2}} #3}{\partial #4\IfNoValueTF{#2}{}{^{#2}}}}
		{\fractype{\partial^2 #3}{\partial #4 \partial #5}}
	}
}
\DeclareDocumentCommand\pdv{}{\partialderivative} 

\DeclareDocumentCommand\dotproduct{}{\boldsymbol\cdot} 
\DeclareDocumentCommand\vdot{}{\dotproduct} 

\DeclareDocumentCommand\crossproduct{}{\boldsymbol\times} 
\DeclareDocumentCommand\cross{}{\crossproduct} 

\DeclareDocumentCommand\gradient{ g o d() }{ 
	\IfNoValueTF{#1}{
		\IfNoValueTF{#2}{
			\IfNoValueTF{#3}{\vnabla}{\fbraces{\lparen}{\rparen}{\vnabla}{#3}}
			}
			{\fbraces{\lbrack}{\rbrack}{\vnabla}{#2} \IfNoValueTF{#3}{}{(#3)}}
		}
		{\vnabla #1 \IfNoValueTF{#2}{}{[#2]} \IfNoValueTF{#3}{}{(#3)}}
	}

\DeclareDocumentCommand\divergence{ g o d() }{ 
	\IfNoValueTF{#1}{
		\IfNoValueTF{#2}{
			\IfNoValueTF{#3}{\vnabla \vdot}{\vnabla \vdot \quantity(#3)}
			}
			{\vnabla \vdot \quantity[#2] \IfNoValueTF{#3}{}{(#3)}}
		}
		{\vnabla \vdot #1 \IfNoValueTF{#2}{}{[#2]} \IfNoValueTF{#3}{}{(#3)}}
	}

\DeclareDocumentCommand\curl{ g o d() }{ 
	\IfNoValueTF{#1}{
		\IfNoValueTF{#2}{
			\IfNoValueTF{#3}{\vnabla \cross}{\vnabla \cross \quantity(#3)}
			}
			{\vnabla \cross \quantity[#2] \IfNoValueTF{#3}{}{(#3)}}
		}
		{\vnabla \cross #1 \IfNoValueTF{#2}{}{[#2]} \IfNoValueTF{#3}{}{(#3)}}
	}

\DeclareDocumentCommand\laplacian{ g o d() }{ 
	\IfNoValueTF{#1}{
		\IfNoValueTF{#2}{
			\IfNoValueTF{#3}{\nabla^2}{\fbraces{\lparen}{\rparen}{\nabla^2}{#3}}
			}
			{\fbraces{\lbrack}{\rbrack}{\nabla^2}{#2} \IfNoValueTF{#3}{}{(#3)}}
		}
		{\nabla^2 #1 \IfNoValueTF{#2}{}{[#2]} \IfNoValueTF{#3}{}{(#3)}}
	}
\begin{document}
\title{Opacity of relativistically underdense plasmas for extremely intense laser pulses}
\author{M.~A.~Serebryakov}
\ead{serebryakovma@ipfran.ru}
\author{A.~S.~Samsonov}
\author{E.~N.~Nerush}
\ead{nerush@ipfran.ru}
\author{I.~Yu.~Kostyukov}
\address{Institute of Applied Physics of the Russian Academy of
Sciences, 46 Ulyanov St., Nizhny Novgorod 603950, Russia}

\begin{abstract}
    It is generally believed that relativistically underdense plasmas is transparent for intense
    laser radiation. However, particle-in-cell simulations reveal abnormal laser field absorption
    above the intensity threshold about~$3 \times 10^{24}~\mathrm{W}\,\mathrm{cm}^{-2}$ for the
    wavelength of $1~\mu \mathrm{m}$. Above the threshold, the further increase of the laser
    intensity doesn't lead to the increase of the propagation distance. The simulations take into
    account emission of hard photons and subsequent pair photoproduction in the laser field. These
    effects lead to onset of a self-sustained electromagnetic cascade and to formation of dense
    electron-positron ($e^+e^-$) plasma right inside the laser field. The plasma absorbs the field
    efficiently, that ensures the plasma opacity. The role of a weak longitudinal electron-ion
    electric field in the cascade growth is discussed.
\end{abstract}

\noindent{\it Keywords\/}:

\maketitle
\ioptwocol
\section{\label{sec:introduction}Introduction}

Propagation of laser radiation through electron-ion plasmas have been studied for decades. For weak
laser field plasmas become opaque if its electron density, $n_e$, exceeds the critical density,
$n_{cr} = m \omega^2 / 4 \pi e^2$, with $\omega = 2 \pi c / \lambda$ the laser cyclic frequency and
$\lambda$ the laser wavelength, $m$ and $e$ are the electron mass and absolute charge value,
respectively. Overdense plasmas, $n_e \gtrsim n_{cr}$, reflects weak laser pulses.  However, if the
laser pulse is relativistically intense, $a_0 \gtrsim 1$, the overdense plasma can become
transparent due to relativistic mass increase of plasma electrons~\cite{Tushentsov01}; here $a_0 =
e E_0 / mc \omega$ is the dimensionless amplitude, $E_0$ is the amplitude of the laser electric
field. In this case --- the case of relativistic self-induced transparency (RSIT) --- a circularly
polarized laser radiation can propagate in dense plasmas ($n_e \gtrsim n_{cr}$) over a finite
length which increases with the incident intensity~\cite{Tushentsov01, Eremin10}.  RSIT have been
investigated in simulations and experiments including the case of linear
polarization~\cite{Palaniyappan12, Weng12b}.

The criterion for RSIT is that the effective plasma density is lower than the plasma critical
density, $n_e / \gamma \lesssim n_{cr}$, with $\gamma$ the electron Lorentz factor. Plasma
electrons are pushed by the laser pulse hence $n_e$ can differ significantly from the initial
plasma density, $n_0$.  Heating and acceleration of the electrons depend on the laser polarization,
on the density gradient at the boundary, etc., however, the Lorentz factor can be roughly estimated
as $a_0$, that corresponds to the energy that a motionless electron gains in the oscillating or
rotating electric field $E_0$. Therefore, RSIT occurs for field intensity
\begin{equation}
    \label{rsit-criterion}
    a_0 \gtrsim \kappa n_0 / n_{cr},
\end{equation}
with $\kappa$ a numeric coefficient of the order of unity. Analytics~\cite{Cattani00} and
simulations~\cite{Lefebvre95, Eremin10, Tushentsov01, Weng12b, Palaniyappan12} show that $\kappa$
is not more than $2$.

The criterion~(\ref{rsit-criterion}) has a simple electrodynamical meaning: in RSIT regime, the
current produced by all the electrons involved in the interaction cannot generate the field which
quenches the incident field inside the plasma, even if the electrons are accelerated up to the
speed of light and the skin layer width is about the wavelength~\cite{Nerush14}.  Therefore, nor
ion motion~\cite{Lefebvre95, Tushentsov01} nor radiation reaction~\cite{Brady12, Nerush14} can
change the RSIT criterion significantly. Thus, in the general sense the plasma which density
fulfils criterion~(\ref{rsit-criterion}) is called relativistically underdense for the given laser
light.

The effect of quantum electrodynamics (QED), i.e. pair photoproduction in electromagnetic
cascades~\cite{Bell08, Fedotov10, Elkina11}, can change the plasma density dramatically. Namely
avalanche-like QED cascades can generate electron-positron plasma which density is above the
opacity threshold~(\ref{rsit-criterion}), hence the plasma absorbs the laser radiation efficiently.
This scenario is observed in simulations for counter-propagating laser beams and seeded
cascades~\cite{Nerush11a, Grismayer16}, where the laser beams form a standing wave which is
suitable for the cascades. For laser beams interacting with plasmas for
counter-propagating~\cite{Zhang15} and single~\cite{Nerush15, Wang17, Samsonov19} laser pulses the
QED effects also influence the laser energy absorption and the plasma opacity. However,
avalanche-like QED cascades do not develop in ordinary plane-wave geometry and need quite dense
plasma --- to develop in the sum of the incident and the reflected field, or to develop on the
plasma-vacuum interface without reflection~\cite{Samsonov19}. Alternatively, noticeable transverse
field gradient is needed for cascade development in the field of a single laser
pulse~\cite{Mironov21}.

In relativistically underdense plasma the laser pulse pushes electrons out (as well as ions at high
intensities) thus a channel is formed. This regime differs significantly from the quasi
one-dimensional hole boring or light sail regimes which occur at higher densities, and where QED
processes have been studied recently~\cite{Nerush15, Samsonov19}. The previous simulations of the
channelling regime performed for $a_0 \lesssim 10^3$ revealed that the laser pulse can trap plasma
electron which field drags ions, hence dense electron-ion bunch is formed right in the laser
pulse~\cite{Ji14b, Ji18, Capdessus20}. Radiation reaction plays a crucial role in this electron
trapping and plasma transparency~\cite{Ji18, Capdessus20}, however, the laser pulse still can
travel a long distance despite of the electron trapping and ``snow plough'' effects. Also, pair
photoproduction for such field strength lead to a small number of positrons whose field is not
enough to influence the laser pulse propagation.

In this paper propagation of spatially limited laser pulses of $a_0 = 300 - 3000$ in a plasma
half-space of density $n_0 = 50 n_{cr}$ is investigated. It is demonstrated that at higher
intensities ($a_0 \gtrsim 1000$) vast amount of electrons and positrons can be generated in the QED
cascade. The $e^+e^-$ pairs are generated not at the laser front, but right inside the pulse, and
absorb the laser energy very efficiently. Similar process have been studied in preplasma attached
to a dense plasma slab~\cite{Wang17}, however, the cascade mechanism can be different there. Also,
in~\cite{Wang17} the absorption rate of the laser energy (per preplasma length) decreases with the
increase of the intensity. Contrary, in the half-space of tenuous plasmas the cascade develops
until the laser energy is absorbed, and the distance on which this happens almost does not depend
on the laser intensity.  Thus, a layer of a tenuous plasma of width about $20 \lambda$ becomes an
impassable barrier for extremely dense laser pulses.

\section{\label{sec:simulations}Numerical simulations}

Absorption of extremely strong laser field by relativistically underdense plasmas is studied with
three-dimensional particle-in-cell (PIC) code QUILL~\cite{Quill}. Besides plasma effects, the code
simulates emission of hard photons and pair photoproduction, both in local constant field
approximation using quasiclassical formulas~\cite{Baier98}. This approximation is widely used for
ultrarelativistic electrons and positrons in the strong field, $a_0 \gg 1$, for any value of
quantum parameter $\chi$ (i.e. for $\chi < 1$ and $\chi > 1$).

In the simulations the laser pulse with amplitude $a_0$ from $300$ to $3000$ is moving in a $20
\times 12 \times 12~\lambda^3$ box. The box is moving with the speed of light to track the pulse
motion. The pulse is linearly polarized along the $y$ axis, the laser wavelength is
$1~\mu\mathrm{m}$, and the envelope is cosine-like $E_y \propto \cos(\pi x / 2 x_s) \cos(\pi y / 2
y_s) \cos(\pi z / 2 z_s)$ with scales along $x$, $y$ and $z$ axes equal to $x_s \times y_x \times
z_s = 5 \times 4.5 \times 4.5~\lambda^3$. Plasma fills the half-space $x > 10 \lambda$ and its
density is $n_e = 50 n_{cr}$; the ion mass-to-charge ratio is twice of that for the protons.

\begin{figure*}[h!]
    \includegraphics[width=1\linewidth]{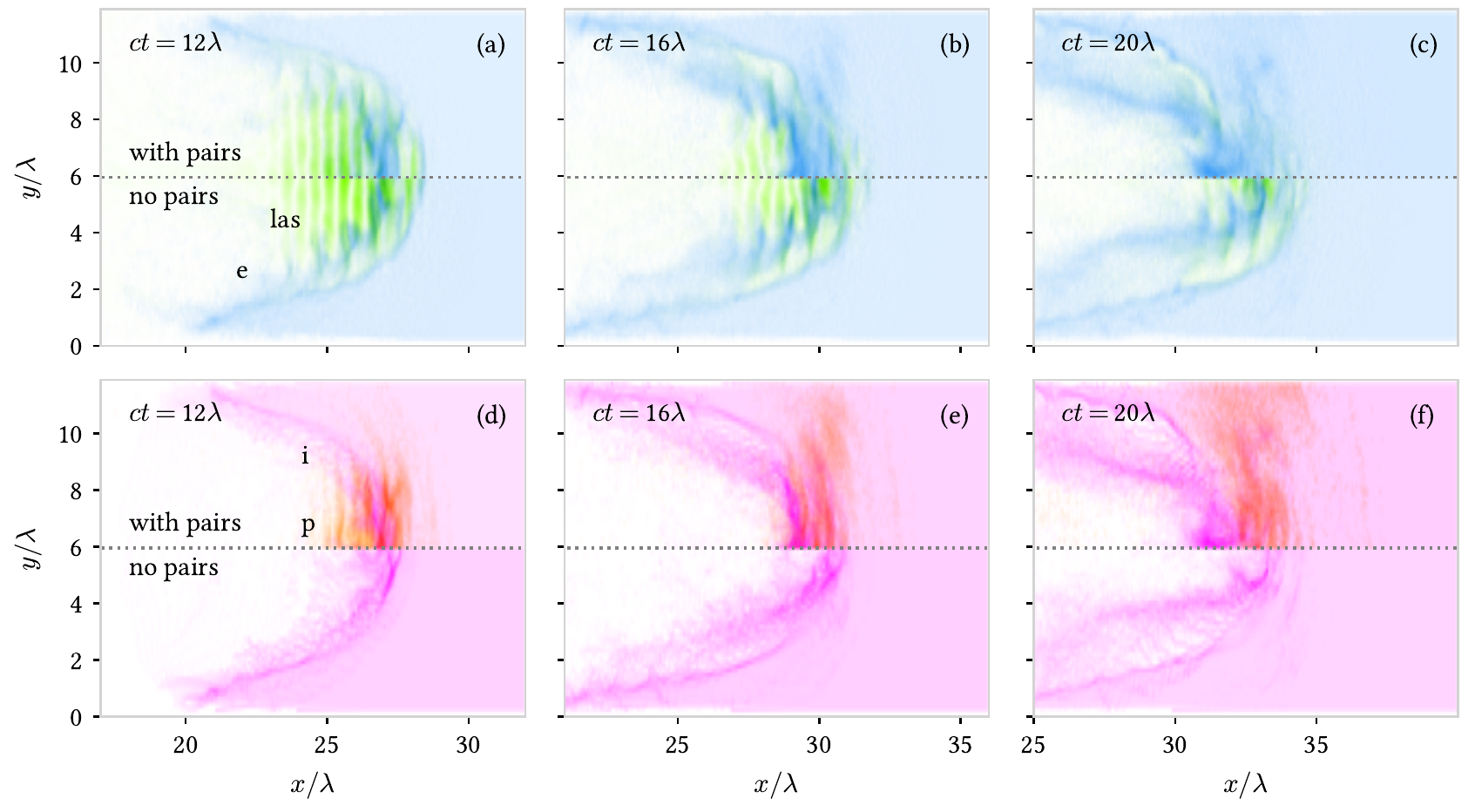}
    \caption{(a)---(f) Particle density and density of electromagnetic energy in the simulations
    for $a_0 = 2100$ and background plasma density $n_e = 50 n_{cr}$. The upper half of each figure
    is the result of modeling with enabled $e^+e^-$ pairs production, and the bottom half is for
    simulations with disabled pair production.  Laser field ``las'' is depicted with light green;
    the electron density ``e'', the positron density ``p'' and the ion density ``i'' are shown in
    blue, orange and magenta, respectively.
    \label{fig:density_comparison}}
\end{figure*}

Two series of simulations have been performed: the first one takes into account both photon
emission and $e^+e^-$ pair production, the second one is non-physical with emission of hard photons
taken into account, but pair photoproduction was artificially turned off. As shown below, the
comparison between these simulations highlights the role of the electron-positron pairs in the
absorption of the laser energy.

The simulation results for $a_0 = 2100$ are shown in Figure~\ref{fig:density_comparison}. The upper
half of every subplot corresponds to the simulation where photoproduction is enabled, and the lower
half --- where photoproduction is disabled. In both cases the laser pulse pushes out the electrons
and the ions. Also, the ions are dragged by the electrons that lead to a hollow channel behind the
laser pulse. A dense electron-ion bunch is presented directly inside the laser pulse [see $x
\approx 31 \lambda$ and $x \approx 33 \lambda$ in upper/lower parts of
\fref{fig:density_comparison}(c)], that is caused by the radiation-reaction trapping~\cite{Ji14b}.
However, in the case of enabled pair photoproduction the electron-ion bunch is much denser and
longer than in the case of disabled photoproduction.

The laser pulse absorption is also different in the cases of on/off pair photoproduction. The
propagation length $L_p$ is introduced as the distance on which the laser pulse loses $50\%$ of its
energy.  The dependence of the propagation length on the laser amplitude $a_0$ is shown in
\fref{fig:energy}(a). Whereas in the case of disabled photoproduction the propagation length grows
linearly with the laser amplitude, in the case of enabled $e^+e^-$ pairs $L_p$ doesn't grow if the
amplitude is above the threshold of about $a_0 \approx 10^3$. Therefore, in the latter case more
than a half of the laser energy is absorbed by a $11 \lambda$ slab of tenuous plasma ($n_e = 50 \,
n_{cr}$) even if the laser amplitude is such high as $a_0 = 3000$.

Figure~\ref{fig:energy}(b) shows the energy redistribution in the simulation for $a_0 = 1800$ and
enabled photoproduction. The dotted black line shows the level of a half of the initial laser
energy and the propagation length. It is seen that the laser energy is converted mostly to the
energy of hard photons, and the energy of electrons, positrons and ions remains quite small.

Figures~\ref{fig:energy}(c) and (d) show how the absorbed energy is distributed between electrons,
positrons, ions and hard photons at the time instance which corresponds to the propagation length,
for different values of the laser amplitude.  For the case of disabled photoproduction
[figure~\ref{fig:energy}(d)] the propagation distance is long and visible part of the energy goes
out of the simulation box (denoted with ``out''). It is seen from the comparison of the
figures~\ref{fig:energy}(c) and (d) that when photoproduction is enabled, the role of the hard
photons is more valuable, and the role of the ions is less noticeable. As will be seen, this is
explained by larger number of the electrons and positrons (due to photoproduction) and smaller
number of the ions involved (due to smaller propagation distance) in the case of enabled pair
photoproduction. These figures do not give a clear hint about the origin of the positrons, however
in the next sections the cascade nature of the pair production is demonstrated, and it is shown
that a single-step process of pair production (e.g. generation from the photons emitted by the
plasma electrons) leads to rather linear dependence $L_p(a_0)$ which does not fit
figure~\ref{fig:energy}(a).

\section{Absorption models}

\begin{figure*}
    \includegraphics{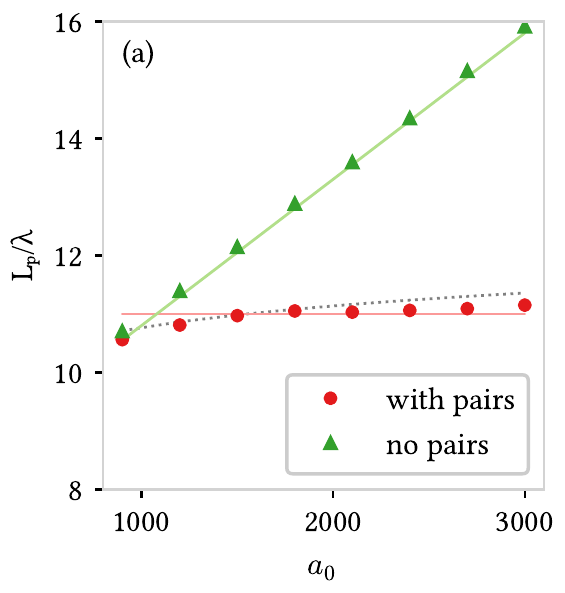}
    \includegraphics{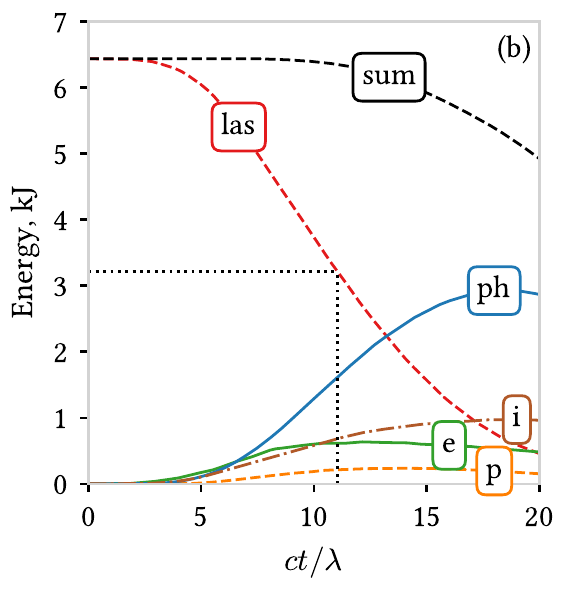}
    \includegraphics{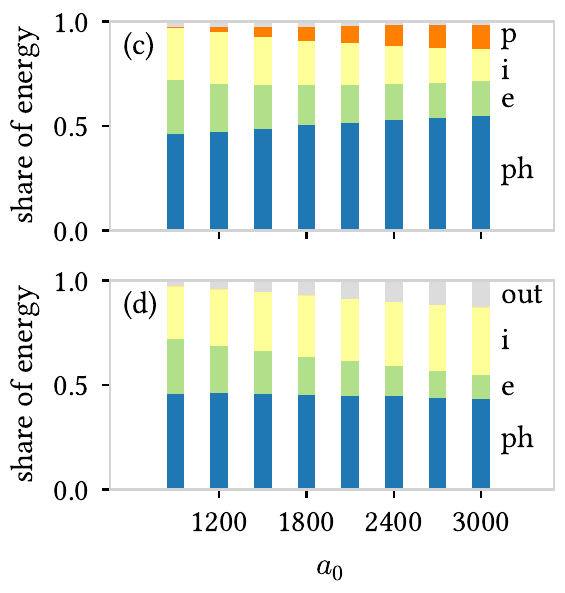}
    \caption{
    (a) The propagation distance of the laser pulse in the relativistically underdense plasma of
    $n_e = 50 n_{cr}$ in numerical simulations with $e^+e^-$ pair photoproduction disabled (green
    triangles) and enabled (red circles). The oblique green solid line and dashed line corresponds
    to absorption models~(\ref{propagation_distance_no_pairs}) and~(\ref{Lppairs}), respectively.
    The horizontal red solid line depicts a constant propagation length.
    (b) Evolution of laser energy ``las'' in the simulation for $a_0 = 1800$ and enabled pair
    photoproduction. The laser energy is partitioned between hard photons ``ph'', electrons ``e'',
    positrons ``p'' and ions ``i''. Black dotted line indicates the point where 50\% of the laser
    energy is absorbed which corresponds to the laser propagation length $L_p$ by the definition.
    (c), (d) Distribution of the absorbed energy at $t = L_p / c$ between different particle kinds
    for enabled and disabled pair photoproduction.}
    \label{fig:energy}
\end{figure*}

In the case of disabled pair photoproduction the propagation distance $L_{p}$ can be estimated as
follows.  The electrons can be considered as the main cause of the dissipation, because they gain
energy efficiently and then convert it to the energy of hard photons which therefore carry away up
to 50\% of the laser energy, see figures~\ref{fig:energy}~(b)---(d). At the same time, the energy
of an electron gained in the laser field during the laser period $\lambda / c$ can be estimated as
$mc^2 a_0$.  This energy is emitted constantly, thus assuming that the electron stays in the field
for about $L_{las} / c$, the emitted energy is $\sim mc^2 a_0 L_{las} / \lambda$. In this case the
energy balance in the system can be described with the following relation:
\begin{equation}
\label{conservation_law_no_pairs}
s n_e mc^2 a_0 L_{las} (L_p - L_i) / \lambda \approx L_{las} \frac{E_0^2}{4\pi}, 
\end{equation}
with $s$ the multiplier which shows how efficiently electrons are accelerated and how long they
stay in the laser field, $L_i$ a distance the laser pulse travels in the simulations before a
steady propagation in the plasma begins, $L_{las}$ the laser pulse length. Thus,
\begin{equation}
\label{propagation_distance_no_pairs}
\frac{L_p - L_i}{\lambda} \approx \frac{a_0 n_{cr}}{s n_e}. \label{propagation_distance_no_pairs}
\end{equation}
To match the simulation results [triangles in figure~\ref{fig:energy}~(a)], one should choose $L_i
= 8.3 \lambda$ and $s = 8$ in equation~(\ref{propagation_distance_no_pairs}). Such large value of
$s$ can be explained with the ``plow effect'' and the radiation reaction trapping which forces the
electrons to stay longer in the laser field. Despite of its approximate nature, the
model~(\ref{propagation_distance_no_pairs}) gives a linear dependence of the propagation distance
$L_p$ on the laser amplitude $a_0$ [oblique solid green line in figure~\ref{fig:energy}(a)] that is
in good agreement with the simulation results for disabled pair photoproduction.

If one adds a single step of pair production to the model, it does not change the
estimate~(\ref{propagation_distance_no_pairs}) much, because it only changes the multiplier $s$.
Contrary, in the cascade the number of electrons and positrons grows exponentially with time, $N
\propto \exp(\Gamma t)$ (with $\Gamma$ the cascade growth rate), that changes the estimate for the
propagation length dramatically.

The number of particles in the cascade doubles up during time $\sim 1/\Gamma$, and if this time is
much shorter than $L_p/c$ from equation~(\ref{propagation_distance_no_pairs}), the density of the
electron-positron plasma can reach the critical density and absorb the laser
pulse~\cite{Nerush11a}. The energy balance can be written under the assumption that the number of
seeding electrons is about the number of plasma electrons scattered by the laser pulse ($\propto
L_p - L_i$), and the energy emitted per electron is about $mc^2 a_0$, therefore
\begin{equation}
    n_e mc^2 a_0 (L_p - L_i) e^{\Gamma (L_p - L_i) / c} =
    L_{las} \frac{E_0^2}{4\pi},
\end{equation}
with $L_i/c$ the time of initial laser-plasma interaction before the cascade starts. Thus $L_p$ can
be found as the solution of the equation
\begin{equation}
    \label{Lppairs}
    L_p - L_i = \frac{c}{\Gamma} \log \left( \frac{L_{las}}{L_p - L_i} \frac{a_0 n_{cr}}{n_e}
    \right),
\end{equation}
which is plotted in \fref{fig:energy}~(a) as dark dashed line, for $\Gamma = 1.5$, $\L_i = 8.3
\lambda$ (as for the case of disabled pair production) and $L_{las} = x_s = 5 \lambda$.  Such value
of the cascade growth rate $\Gamma$ looks reasonable: according to the
simulations~\cite{Grismayer2017}, for counter-propagating laser pulses $\Gamma \approx 2$ for $a_0
\approx 2000$.  Furthermore, for $L_p - L_i \sim L_{las}$ the logarithm function in
equation~(\ref{Lppairs}) changes so slowly with $a_0$ (because $a_0 n_{cr} \gg n_e$) that the
propagation distance can be considered as a constant which do not depend on $a_0$ at all, $L_p
\approx 11 \lambda$ [see horizontal solid line in \fref{fig:energy}~(a)].

\begin{figure*}[htbp]
    \includegraphics{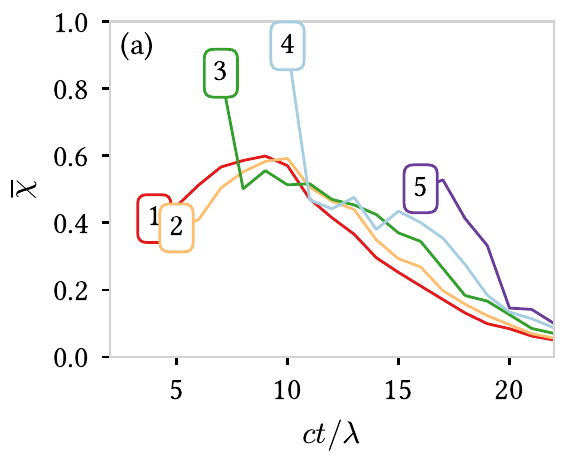}
    \includegraphics{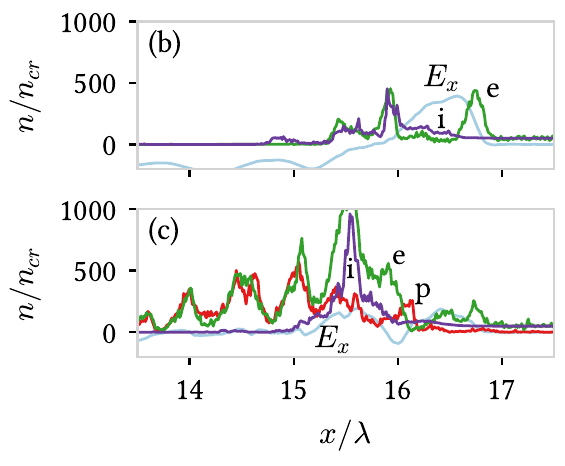}
    \includegraphics{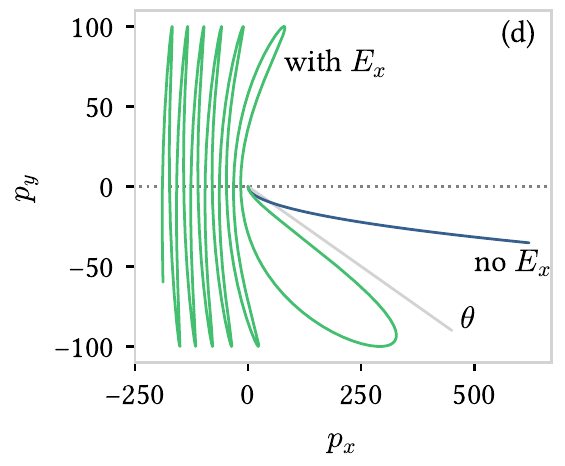}
    \caption{\label{fig:generations}
    (a) Average $\chi$ parameter of positrons of different generations.
    (b) Longitudinal electric field $E_x$, ion ``i'' and electron ``e'' density along the
    longitudinal axis of the laser pulse, for $a_0 = 2500$, $t = 11 \lambda / c$ and disabled
    photoproduction.
    (c) The same as in (b), but for enabled pair photoproduction, ``p'' labels the positron
    density.
    (d) The electron trajectories in the field of a plane wave, with and without additional
    longitudinal decelerating electric field. See text for further details.}
\end{figure*}

In the additional simulation with $a_0=2500$ we track particle generations. The initial plasma
electrons are labelled as the zeroth generation.  Plasma electrons emit photons which produce
electrons and positrons of the first generation.  These electrons and positrons produce photons
which then can produce the second generation of the electrons and positrons, and so on.
Figure~\ref{fig:generations}~(a) depicts the average $\chi$ parameter of the positrons of every
generation, $\bar \chi$. There are only five generations of particles, and $\bar \chi$ value is
below 1, but it is enough to produce dense electron-positron plasma.  Electron ``e'', ion ``i'' and
positron ``p'' density, as well as the longitudinal electric field $E_x$ along the laser pulse axis
are shown in figures~\ref{fig:generations}~(b) and (c), which correspond to the cases of disabled
and enabled pair photoproduction, respectively. Thus, the simulations clearly demonstrate the
cascade nature of the laser absorption, and in the next section we analyze how electrons can gain
$\chi$ parameter in the considered field configuration.

\section{Electron motion in the plane wave and weak decelerating field \label{motion_pv}}

It can be shown that in a pure plane wave field the quantum parameter $\chi$ of an electron (or a
positron) does not increase~\cite{Samsonov19}. Also, the emission of synchrotron photons makes the
parameter $\chi$ smaller. Therefore, the self-sustained QED cascades~\cite{Bell08, Fedotov10} can't
develop in the plane-wave field, and the origin of the QED cascade in the model of the anomalous
absorption in the previous Section should be explained. We claim that a small decelerating
longitudinal plasma field (itself not sufficient to support a cascade) is enough to change the
electron motion dramatically and lead to the increase of $\chi$.

Let's consider the electron motion in the field of a linearly polarized plane wave which propagates
along the $x$ axis (with non-zero field components $E_y = B_z$) and a constant longitudinal field
$E_x > 0$.  It is convenient to describe the plane-wave field with the vector potential $\vb{A} =
\vu{y} A_y(x - t)$, with $\vu{y}$ the unit vector along the $y$ axis. For the sake of simplicity,
the electron with zero momentum along the $z$ axis is considered, $p_z = 0$, and for the rest of
the components one has
\begin{eqnarray}
    \dv{\vb{p}}{t} = - \vb{E} - \vb{v} \times \vb{B}, \\
    \dv{p_x}{t} = - v_y \pdv{A_y}{x} - E_x, \label{force} \\
    \dv{p_y}{t} = \pdv{A_y}{t} + v_x \pdv{A_y}{x} \label{dot_py},
\end{eqnarray}
with $\vb{p}$ the electron momentum normalized to $mc$, $E$ and $B$ the electric and magnetic
fields normalized to $mc\omega/e$, $t$ the time normalized to $1/\omega$ and $\vb{v} = \vb{p} /
\gamma$ the electron velocity.  In the case of pure plane wave field ($E_x = 0$) two integrals of
motion are known, $\gamma - p_x$ and $p_y - A_y$. For non-zero longitudinal field the latter still
conserves,
\begin{equation}
p_y - A_y = C_y, \label{py}
\end{equation}
and for the former with the equation for the Lorentz factor, $d\gamma / dt = - (v_x E_x + v_y
E_y)$, one gets
\begin{eqnarray}
    \frac{d(\gamma - p_x)}{dt} = (1 - v_x) E_x, \\
    \gamma - p_x = C_x - \xi \, E_x . \label{px}
\end{eqnarray}
Here $C_{x,y}$ are constants and $\xi = x - t$.

With equation~(\ref{py}) the dependence $p_y(\xi) = C_y + A_y(\xi)$ is easily found, then $p_x$ is
found from equation~(\ref{px}) as follows:
\begin{equation}
    p_x(\xi) = \frac{1 + [C_y + A_y(\xi)]^2 - [C_x - \xi E_x]^2}{2 (C_x - \xi E_x)}.
\end{equation}
With known $p_y(\xi)$ and $p_x(\xi)$, the dependence $t(\xi)$ is found from the numerical
integration of $dt/d\xi = 1/(v_x - 1)$. The resulting electron trajectory for a plane wave with
amplitude $a_0 = 100$ and $E_x = 10$ is shown in Fig.~\ref{fig:generations}(d) with the light green
line. The dark blue line corresponds to the electron trajectory in the case of zero longitudinal
field. Initially, at $t = 0$, the electron is motionless, $p_x = p_y = 0$, and the trajectories are
shown up to $t / (2 \pi) = 12$. In the case of zero $E_x$ the electron trajectory in the momentum
space is the parabola $p_x = p_y^2/2$, and the longitudinal velocity is very close to unity for
$a_0 \gg 1$: $p_y \sim a_0$, $p_x \gg p_y$. Therefore, the phase $\xi$ changes very slowly, and the
electron moves with the wave most of time. In the case of non-zero $E_x$ the longitudinal velocity
is not so close to the speed of light hence the phase $\xi$ changes much faster, and the electron
doesn't move with the wave. The clue to this dramatic change in the electron motion is seen already
in the crossed fields.

Let's consider the electron motion in the constant crossed fields $E_y = B_z$ with weak
decelerating field $E_x \ll |E_y|$. Without $E_x$, the initially motionless electron is accelerated
to $v_x \approx 1$, but non-zero $E_x > 0$ makes motion along the $x$ axis unstable. The new stable
direction can be found as an equilibrium state, i.e. the Lorentz force transverse to this direction
should be equal to zero. Thus, assuming $v \approx 1$, the angle $\theta$ between the $x$ axis and
this direction (measured clockwise) is
\begin{eqnarray}
    E_y \cos \theta + E_x \sin \theta = B_z, \\
    \theta \approx 2 E_x / E_y. \label{theta}
\end{eqnarray}

Returning to the plane wave, one can note that motion not along the $x$ axis but along the stable
direction corresponding to equation~(\ref{theta}) coerces the phase to change faster. The sign of
$E_y$ changes with the phase, hence the sign of $\theta$ also changes, which makes the electron to
move transverse to the wave and even towards the wave when it moves between the directions
corresponding to the wave amplitude $\theta = \pm 2 a_0 / E_x$ (one of them is shown in
Fig.~\ref{fig:generations}(d) with light gray line).  In this case the quantum parameter $\chi$ is
much greater than in the case of the electron motion along the $x$ axis, in which the Lorentz force
is minimal. Therefore small decelerating field added to a plane wave can make the QED cascade
possible.

\section{Conclusions}

In conclusion, we have shown that the QED cascade process is crucial for the absorption of an
extremely intense laser pulse by relativistically underdense plasmas. Namely, QED-PIC simulations
show that laser pulses lose a half of their energy on a distance of about $10~\mu\mathrm{m}$ in
plasma of density $50 \, n_{cr}$,  for the laser amplitude above the threshold $a_0 = 1 \times
10^3$ and up to $a_0 = 3 \times 10^3$  (for $\lambda = 1~\mu\mathrm{ m}$). Contrary to the
simulations with $e^+e^-$ pair photoproduction taken into account, the propagation distance
increases with the increase of the laser intensity if only photon emission but not pair
photoproduction is taken into account.  Two simple models are proposed for laser absorption in the
cases of disabled/enabled pair photoproduction in the simulations. The models describe fairly well
the dependence of the propagation distance on the laser amplitude.

Although QED cascades have been considered mostly in dense plasmas or in setups with multiple laser
pulses, there is a reasoning about the cascade nature of the unexpectedly strong absorption in the
relativistically underdense plasmas. First, beyond the intensity threshold, the density of the
produced electron-positron plasma is much higher than the initial plasma density. Second,
simulations directly demonstrate several generations of the secondary particles, i.e. several steps
of subsequent photon emission and pair photoproduction. Third, radiation reaction
trapping~\cite{Ji14b} leads to formation of dense electron-ion bunch inside the laser pulse. The
bunch density is not enough for efficient absorption, but the bunch generates weak longitudinal
electric field (of the order of $0.1$ of the laser amplitude).  Analysis of the electron motion
shows that such field being added to a field of a plane wave allows the electron to gain the
quantum parameter $\chi$, hence, makes the avalanche-like cascade possible.

The absorption of the laser energy is important for various applications of extremely intense
lasers, such as ion and electron acceleration or generation of hard photons. Also, some
astrophysical conditions assume interaction of extreme electromagnetic fields with underdense
plasmas. Thus, the parametric study of the anomalously strong absorption is extremely interesting
and will be the subject of a future research.

\ack The research is supported by the Russian Science Foundation (Grant No. 20-12-00077).
\\
\\
\bibliographystyle{unsrt}
\bibliography{radiation-trapping}

\begin{thebibliography}{10}

\bibitem{Tushentsov01}
M.~Tushentsov, A.~Kim, F.~Cattani, D.~Anderson, and M.~Lisak.
\newblock Electromagnetic energy penetration in the self-induced transparency
  regime of relativistic laser-plasma interactions.
\newblock {\em Physical Review Letters, vol. 87, Issue 27, id. 275002},
  87:--315002, dec 2001.

\bibitem{Eremin10}
V.~I. Eremin, A.~V. Korzhimanov, and A.~V. Kim.
\newblock Relativistic self-induced transparency effect during ultraintense
  laser interaction with overdense plasmas: Why it occurs and its use for
  ultrashort electron bunch generation.
\newblock {\em Physics of Plasmas}, 17(4):043102, 2010.

\bibitem{Palaniyappan12}
Sasi Palaniyappan, B.~Manuel Hegelich, Hui-Chun Wu, Daniel Jung, Donald~C.
  Gautier, Lin Yin, Brian~J. Albright, Randall~P. Johnson, Tsutomu Shimada,
  Samuel Letzring, Dustin~T. Offermann, Jun Ren, Chengkun Huang, Rainer
  H{\"{o}}rlein, Brendan Dromey, Juan~C. Fernandez, and Rahul~C. Shah.
\newblock Dynamics of relativistic transparency and optical shuttering in
  expanding overdense plasmas.
\newblock {\em Nature Physics}, 8:763--769, oct 2012.

\bibitem{Weng12b}
S.~M. Weng, P.~Mulser, and Z.~M. Sheng.
\newblock Relativistic critical density increase and relaxation and
  high{-}power pulse propagation.
\newblock {\em Physics of Plasmas}, 19(2):022705, 2012.

\bibitem{Cattani00}
F.~Cattani, A.~Kim, D.~Anderson, and M.~Lisak.
\newblock Threshold of induced transparency in the relativistic interaction of
  an electromagnetic wave with overdense plasmas.
\newblock {\em Phys. Rev. E}, 62(1):1234--1237, Jul 2000.

\bibitem{Lefebvre95}
Erik Lefebvre and Guy Bonnaud.
\newblock Transparency/opacity of a solid target illuminated by an
  ultrahigh-intensity laser pulse.
\newblock {\em Phys. Rev. Lett.}, 74(11):2002--2005, Mar 1995.

\bibitem{Nerush14}
E.~N. Nerush, I.~Yu. Kostyukov, L.~Ji, and A.~Pukhov.
\newblock Gamma-ray generation in ultrahigh-intensity laser-foil interactions.
\newblock {\em Physics of Plasmas}, 21:013109, 2014.

\bibitem{Brady12}
C.~S. Brady, C.~P. Ridgers, T.~D. Arber, A.~R. Bell, and J.~G. Kirk.
\newblock Laser absorption in relativistically underdense plasmas by
  synchrotron radiation.
\newblock {\em Phys. Rev. Lett.}, 109(24):245006, Dec 2012.

\bibitem{Bell08}
A.~Bell and John Kirk.
\newblock Possibility of prolific pair production with high-power lasers.
\newblock {\em Physical Review Letters}, 101(20):200403, 2008.

\bibitem{Fedotov10}
A.~M. Fedotov, N.~B. Narozhny, G.~Mourou, and G.~Korn.
\newblock Limitations on the attainable intensity of high power lasers.
\newblock {\em Phys. Rev. Lett.}, 105(8):080402, Aug 2010.

\bibitem{Elkina11}
N.~V. Elkina, A.~M. Fedotov, I.~Yu. Kostyukov, M.~V. Legkov, N.~B. Narozhny,
  E.~N. Nerush, and H.~Ruhl.
\newblock Qed cascades induced by circularly polarized laser fields.
\newblock {\em Physical Review Special Topics - Accelerators and Beams},
  14(5):054401, may 2011.

\bibitem{Nerush11a}
E.~N. Nerush, I.~Yu. Kostyukov, A.~M. Fedotov, N.~B. Narozhny, N.~V. Elkina,
  and H.~Ruhl.
\newblock Laser field absorption in self-generated electron-positron pair
  plasma.
\newblock {\em Phys. Rev. Lett.}, 106(3):035001, Jan 2011.

\bibitem{Grismayer16}
T.~Grismayer, M.~Vranic, J.~L. Martins, R.~A. Fonseca, and L.~O. Silva.
\newblock Laser absorption via quantum electrodynamics cascades in counter
  propagating laser pulses.
\newblock {\em Physics of Plasmas}, 23(5):056706, may 2016.

\bibitem{Zhang15}
Peng Zhang, C.~P. Ridgers, and A.~G.~R. Thomas.
\newblock The effect of nonlinear quantum electrodynamics on relativistic
  transparency and laser absorption in ultra-relativistic plasmas.
\newblock {\em New Journal of Physics}, 17(4):043051, 2015.

\bibitem{Nerush15}
E.~N. Nerush and I.~Y. Kostyukov.
\newblock Laser-driven hole boring and gamma-ray emission in high-density
  plasmas.
\newblock {\em Plasma Physics and Controlled Fusion}, 57(3):035007, 2015.

\bibitem{Wang17}
W.-M. Wang, P.~Gibbon, Z.-M. Sheng, Y.-T. Li, and J.~Zhang.
\newblock Laser opacity in underdense preplasma of solid targets due to quantum
  electrodynamics effects.
\newblock {\em Physical Review E, Volume 96, Issue 1, id.013201}, 96:013201,
  jul 2017.

\bibitem{Samsonov19}
A.~S. Samsonov, E.~N. Nerush, and I.~Yu Kostyukov.
\newblock Laser-driven vacuum breakdown waves.
\newblock {\em Scientific reports}, 9:11133, 2019.

\bibitem{Mironov21}
A.~A. Mironov, E.~G. Gelfer, and A.~M. Fedotov.
\newblock Onset of electron-seeded cascades in generic electromagnetic fields.
\newblock {\em eprint arXiv:2105.04476}, 2105:arXiv:2105.04476, may 2021.

\bibitem{Ji14b}
L.~L. Ji, A.~Pukhov, I.~Yu. Kostyukov, B.~F. Shen, and K.~Akli.
\newblock Radiation-reaction trapping of electrons in extreme laser fields.
\newblock {\em Phys. Rev. Lett.}, 112(14):145003, Apr 2014.

\bibitem{Ji18}
Liangliang Ji, Baifei Shen, and Xiaomei Zhang.
\newblock Transparency of near-critical density plasmas under extreme laser
  intensities.
\newblock {\em New Journal of Physics}, 20(5):053043, may 2018.

\bibitem{Capdessus20}
R.~Capdessus, L.~Gremillet, and P.~McKenna.
\newblock High-density electron{--}ion bunch formation and multi-gev positron
  production via radiative trapping in extreme-intensity laser{--}plasma
  interactions.
\newblock {\em New Journal of Physics}, 22(11):113003, nov 2020.

\bibitem{Quill}
QUILL, \url{https://github.com/QUILL-PIC/Quill}.

\bibitem{Baier98}
V.~N. Baier, V.~M. Katkov, and V.~M. Strakhovenko.
\newblock {\em Electromagnetic processes at high energies in oriented single
  crystals}.
\newblock World Scientific, 1998.

\bibitem{Grismayer2017}
T.~Grismayer, M.~Vranic, J.~L. Martins, R.~A. Fonseca, and L.~O. Silva.
\newblock Seeded {QED} cascades in counterpropagating laser pulses.
\newblock {\em Physical Review E}, 95(2):023210, feb 2017.

\end{thebibliography}

\end{document}